\documentclass[conference, 10pt]{IEEEtran}
\IEEEoverridecommandlockouts
\usepackage{cite}
\usepackage{amsmath,amssymb,amsfonts, bbold}
\usepackage{algorithm}
\usepackage{algorithmic}
\usepackage{graphicx}
\usepackage[font=footnotesize]{caption}
\usepackage[font=footnotesize]{subcaption}
\usepackage{textcomp}
\usepackage{xcolor}
\usepackage{comment}

\usepackage{pgfplots}
\pgfplotsset{compat=1.16}
\usepgfplotslibrary{fillbetween}
\pgfplotsset{yticklabel style={text width=1.2em,align=right}}
\definecolor{matblue}{rgb}{0, 0.4470, 0.7410}
\definecolor{matred}{rgb}{0.85, 0.3250, 0.0980}
\definecolor{matorange}{rgb}{0.9290, 0.6940, 0.1250}

\newcommand\numberthis{\addtocounter{equation}{1}\tag{\theequation}}

\newtheorem{theorem}{Theorem}

\begin{document}

\title{Robust and Reliable Stochastic\\ Resource Allocation via Tail Waterfilling \thanks{This work is supported by the NSF under grant CCF 2242215.}
}

\author{Gokberk Yaylali and Dionysis Kalogerias \\ Department of EE -- Yale University }

\maketitle

\begin{abstract}
Stochastic allocation of resources in the context of wireless systems ultimately demands reactive decision making for meaningfully optimizing network-wide random utilities, while respecting certain resource constraints. Standard ergodic-optimal policies are however susceptible to the statistical variability of fading, often leading to systems which are severely unreliable and spectrally wasteful. On the flip side, minimax/outage-optimal policies are too pessimistic and often hard to determine. We propose a new \textit{risk-aware} formulation of the resource allocation problem for standard multi-user point-to-point power-constrained communication with no cross-interference, by employing the Conditional Value-at-Risk (CV@R) as a measure of fading risk. A remarkable feature of this approach is that it is a convex generalization of the ergodic setting while inducing robustness and reliability in a fully tunable way, thus bridging the gap between the (naive) ergodic and (conservative) minimax approaches. We provide a closed-form expression for the CV@R-optimal policy given primal/dual variables, extending the classical stochastic waterfilling policy. We then develop a primal-dual \textit{tail-waterfilling} scheme to recursively learn a globally optimal risk-aware policy. The effectiveness of the approach is verified via detailed simulations.
\vspace{2pt}
\end{abstract}

\begin{IEEEkeywords}
Resource Allocation, Waterfilling, Conditional Value-at-Risk (CV@R), Risk-Aware Optimization.
\end{IEEEkeywords}

\section{Introduction}
\label{sec:introduction}
\vspace{-4pt}
We revisit the classical problem of allocating resources in point-to-point communication networks operating over realizations of random fading channels $\boldsymbol{h} \in \mathcal{H} \subseteq \mathbb{R}^n$. Resources such as transmission power and/or channel access are allotted among users to meaningfully optimize certain network-wide random utilities. Traditionally, such resources are allocated either \textit{deterministically} by essentially disregarding the statistical variability of fading as an integral characteristic of the system, including minimax formulations  \cite{Parsaeefard2013, Mokari2016}, or \textit{stochastically} in an ergodic sense by considering performance averages \cite{Ribeiro2012, Mokari2016, AliHemmati2018, Kalogerias2020, Hashmi2021}, i.e., expectations of random network objectives in an attempt to optimize performance in the ``long-term". 

However, being optimal in expectation, ergodic stochastic resource allocation lacks the ability to effectively quantify relatively infrequent though statistically significant fading events causing performance drops, e.g., deep(er) fades. Indeed, the statistical dispersion of a communication medium with a fatter-tailed distribution is quite likely to result in rather undesirable channel realizations, leading to potentially major service losses. This happens because expectations of random services do not capture such  risky tail events. In other words, ergodic-optimal resource allocation policies are \textit{risk-neutral}. 

In fact, it is well-known that optimal ergodic policies are often channel-opportunistic \cite{Ribeiro2012}, and prone to sporadic channel realizations that negatively affect performance, leading to unreliable systems suffering from substantial spectrum underutilization. On the other extent, minimax-type (sometimes called ``robust") resource policies aim for maximally reliable system performance \cite{Mokari2016, Parsaeefard2013}. Still, such policies are known for being overcautious and for achieving conservative system performance, on top of the often unreasonable difficulty of the resulting optimization problems.

While approaches based on outage probability optimization
try to bridge the two extremes \cite{Li2005mac}, they exhibit counterfactual issues: Outage probability targets required for performing allocation of resources might not even be feasible to begin with, and even if they are, they might not result in operationally meaningful performance.
Quantile-based resource allocation, such as outage rate/capacity optimization, aims for alleviating those issues, however the resulting problems still suffer from other limitations, mainly related to interpretability and lack of favorable structure (e.g., convexity). 


In this paper, we introduce a \textit{risk-aware}  formulation of the resource allocation problem --such ideas/approaches have recently started getting traction \cite{Vu2018, Li2021, Bennis2018}-- for standard multi-user point-to-point resource-constrained communication with no cross-interference, by capitalizing on the \textit{Conditional Value-at-Risk (CV@R)} \cite{Rockafellar2000} as a measure of fading risk. CV@R, deeply rooted in mathematical finance, is a \textit{coherent risk measure} \cite{Shapiro2014}, trades naturally between the (naive) ergodic and (conservative) minimax settings, and allows formulating the proposed risk-aware problem as a convex, well-structured extension of its ergodic counterpart, liberated from counterfactual issues and inducing robustness and reliability in a fully tunable way.

After obtaining a closed-form expression for the optimal Lagrangian-relaxed CV@R policy, we propose the \textit{tail waterfilling algorithm}, a primal-dual scheme to learn a globally optimal risk-aware policy in a recursive fashion. Indeed, tail waterfilling continuously extends classical stochastic waterfilling \cite{Cover2005} to the risk-aware universe. We present detailed numerical simulations, empirically corroborating the effectiveness of our approach for two standard utilities, namely, weighted sumrate and proportional fairness.





\section{System Model}
\label{sec:system_model}

\begin{figure}[htbp]
    \centering
    \includegraphics[width=.15\textwidth]{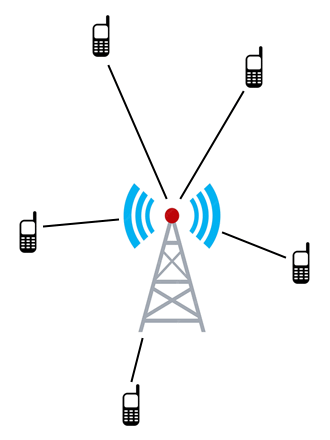}
    \hspace{1cm}
    \includegraphics[width=.15\textwidth]{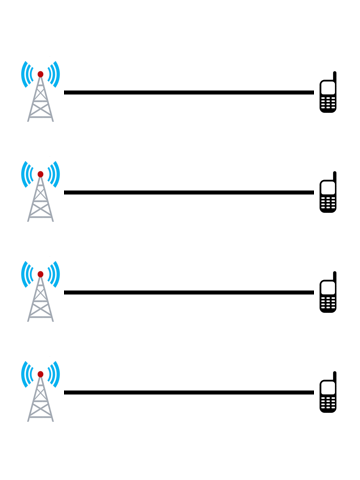}
    \caption{Examples of multi-user one-to-one communication channels. Left: Multiplexed (e.g., time or frequency) star uplink or downlink model. Right: Classical parallel channel model.}
    \label{fig:simo}
\end{figure}

We consider a $n$-terminal parallel point-to-point communication channel model; some examples of relevant networking scenarios may be visualized as in Fig.~\ref{fig:simo}.
We assume perfect channel state information (CSI) at transmission time (mainly for simplicity), which is leveraged to allocate resources via a \textit{policy} $\boldsymbol{p}(\boldsymbol{h})$, where $\boldsymbol{h}$ is the channel fading vector. The rate for terminal or user $i=1,\ldots,n$ in the network is
\begin{equation}
r_i(p_i(h_i),h_i) \triangleq \log \left( 1 + \frac{h_i p_i(h_i)}{\sigma_i^2} \right),
\end{equation}
where $\sigma_i^2>0$ is the noise variance of the corresponding link. Under this setting, optimal resource allocation in an ergodic sense may be achieved by solving the convex problem \cite{Ribeiro2012}
\begin{equation}
\begin{aligned}
\underset{\boldsymbol{x}\in \mathcal{X}, \boldsymbol{p}\succeq \boldsymbol{0}}{\mathrm{maximize}} \quad& f_0 (\boldsymbol{x} ) \\
\mathrm{subject\ to} \quad& \boldsymbol{x} \preceq \mathbb{E} \left[\boldsymbol{r}(\boldsymbol{p}(\boldsymbol{h}), \boldsymbol{h}) \right]\\
& \left\Vert \mathbb{E}\left[ \boldsymbol{p} (\boldsymbol{h}) \right] \right\Vert_1 \leq P_0
\end{aligned}\, ,
\label{eq:primal_riskneutral}
\end{equation}
where $f_0$ is a given concave utility, $\boldsymbol{x}$ is the mean-ergodic rate vector, $\mathcal{X}$ is a convex set, $\boldsymbol{r}$ is the instantaneous rate vector, and $P_0$ is a total mean power budget. Problem \eqref{eq:primal_riskneutral} is very well-studied; in fact, a globally optimal solution may be obtained via the well-known (stochastic) \textit{waterfilling algorithm} \cite{Cover2005}, which is the same as (stochastic) dual descent \cite{Ribeiro2012}. However, as mentioned in Section \ref{sec:introduction}, it is also known that optimal policies obtained by solving \eqref{eq:primal_riskneutral} are channel-opportunistic, unavoidably leading to systems which are severely unreliable and spectrally wasteful. This is because of the \textit{risk-neutral} quantification of channel uncertainty in \eqref{eq:primal_riskneutral}, which discards information about the higher-order or \textit{tail} behavior of the rate as a function of random fading. On the other hand, minimax-optimal policies or policies minimizing outage probabilities are either overly pessimistic \cite{Li2005mac}, or result in problems that are difficult to handle, or suffer from counterfactual issues.

\begin{figure}[!t]
    \centering
    \begin{tikzpicture}
    \begin{axis}[
        width=\linewidth,
        height=.5\linewidth,
        xlabel={Value (of $z$)}, ylabel={Likelihood}, xmin=0, xmax=5,
        grid]
    \addplot[name path = A, blue, thick, domain=0:1.34, samples=500] {x*exp(-x^2 / 2)};
    \addplot[name path = B, red, thick, domain=1.01:5, samples=500] {x*exp(-x^2 / 2)};
    \addplot[name path = C, color=blue!50!red, thick, domain=1.01:1.34, samples=500] {x*exp(-x^2 / 2)};
    \draw[color=red!20!blue, very thick] (axis cs:{1.34,0}) -- (axis cs:{1.34,0.5460});
    \draw[color=red!80!blue, very thick] (axis cs:{1.01,0}) -- (axis cs:{1.01,0.6065});
    \draw[->, blue, very thick] (axis cs:{0.81,0}) -- (axis cs:{0.81,0.08});
    \draw[->, red, very thick] (axis cs:{1.66,0}) -- (axis cs:{1.66,0.08});
    \addplot[name path = D, black, opacity=0, domain=0:5, samples=500] {0};
    \addplot[blue, opacity=0.25] fill between[of=A and D, soft clip={domain=0:1.34}];
    \addplot[red, opacity=0.25] fill between[of=B and D, soft clip={domain=1.01:5}];
    \node[anchor=west] at (axis cs: 1.5,.57) {$\text{area} = \alpha$};
    \node[anchor=west] at (axis cs: 2,.35) {$\text{area} = \alpha$};
    \draw[->, blue, thick] (axis cs:{1.5,0.57}) -- (axis cs:{0.9,0.57});
    \draw[->, red, thick] (axis cs:{2,0.35}) -- (axis cs:{1.6,0.35});
    \node[anchor=south] at (axis cs: 0.81,0.08) {$R_1$};
    \node[anchor=south] at (axis cs: 1.66,0.08) {$R_2$};
    \end{axis}
    \end{tikzpicture}
    \caption{$R_1=-\text{CV@R}^\alpha[-z]$ (blue) and $R_2=\text{CV@R}^\alpha[z]$ (red) calculated at level $\alpha = 0.6$, for a Rayleigh distributed $z$ with scale $\sigma = 1$.}
    \label{fig:cvar_visual}
\end{figure}
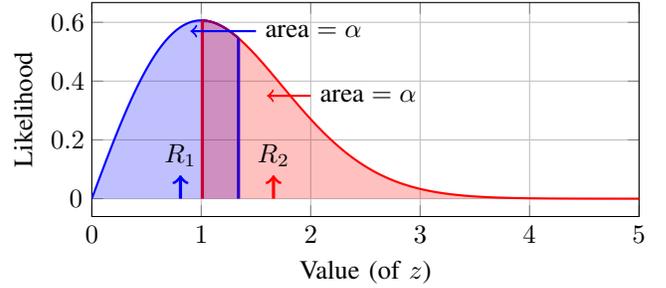

To effectively address those shortcomings, we take a fundamentally distinct approach to stochastic resource allocation by replacing the expectation of rates in \eqref{eq:primal_riskneutral} by a (vector) \textit{risk measure} \cite{Shapiro2014}, specifically the CV@R \cite{Rockafellar2000}, defined for an integrable random \textit{cost} $z$ as
\begin{equation}
\text{CV@R}^\alpha [z] \triangleq \inf_{t \in \mathbb{R}}\ t + \frac{1}{\alpha} \mathbb{E}[(z-t)_+],
\end{equation}
where $\alpha\in (0,1]$ is the corresponding \textit{confidence level}. CV@R is a strict and tractable generalization of expectation, because 
\begin{equation}
    \begin{aligned}
        \text{CV@R}^1 [z] & = \mathbb{E}[z]
        \le \text{CV@R}^\alpha [z], \,\, \forall \alpha \in (0,1] \,\,\text{and} 
        \\
        \text{CV@R}^0 [z] & \triangleq \lim_{\alpha \downarrow 0} \text{CV@R}^\alpha [z] = \text{ess} \hspace{1pt}\text{sup} \, z.
    \end{aligned}
\end{equation}
Intuitively, CV@R measures \textit{expected losses restricted to the upper tail} of $z$ of probability equal to $\alpha$; see Fig. \ref{fig:cvar_visual}. Therefore, it provides an interpretable and tunable tradeoff bridging risk-neutrality and minimax robustness.

To make CV@R suitable for maximizing rewards --cf. \eqref{eq:primal_riskneutral}-- rather than minimizing losses, it is sufficient to reflect it as
\begin{equation}\label{CV@R_sup}
-\text{CV@R}^\alpha [-z] = \sup_{t \in \mathbb{R}}\ t - \frac{1}{\alpha} \mathbb{E}[(t-z)_+],
\end{equation}
now measuring expected \textit{rewards} restricted to the \textit{lower} tail of $z$ of probability equal to $\alpha$; again, see Fig. \ref{fig:cvar_visual} for a comparison.
Using \eqref{CV@R_sup}, we may formulate our proposed risk-aware resource allocation problem as
\begin{equation}
\hspace{4pt}\boxed{\begin{aligned}
P^* = \underset{\boldsymbol{x}\in \mathcal{X}, \boldsymbol{p} \succeq \boldsymbol{0} }{\mathrm{maximize}} \quad& f_0 (\boldsymbol{x} ) \\
\mathrm{subject\ to} \quad& \boldsymbol{x} \preceq - \text{CV@R}^{\boldsymbol{\alpha}} \left[ -\boldsymbol{r}(\boldsymbol{p}(\boldsymbol{h}), \boldsymbol{h}) \right]\\
& \left\Vert \mathbb{E}\left[ \boldsymbol{p} (\boldsymbol{h}) \right] \right\Vert_1 \leq P_0
\end{aligned}}\,\hspace{1pt},
\label{eq:primal_riskaware} 
\end{equation}
where $\boldsymbol{x}$ is now interpreted as a \textit{risk-ergodic rate} vector, and the vector operator $\text{CV@R}^{\boldsymbol{\alpha}}[\cdot]$ with a confidence level vector $\boldsymbol{\alpha}$ evaluates the risk of the corresponding rate vector in an elementwise manner. Note that we have tacitly not enforced risk-aware behavior on the policy itself (resource constraint), as this is operationally unnecessary.
It is a standard exercise to show that problem \eqref{eq:primal_riskaware} can be equivalently expressed as
\begin{equation}
\begin{aligned}
\hspace{-8bp} P^* = \underset{\boldsymbol{x}\in \mathcal{X}, \boldsymbol{p} \succeq \boldsymbol{0}, \boldsymbol{t}}{\mathrm{maximize}} \quad& f_0 (\boldsymbol{x} ) \\
\mathrm{subject\ to} \quad& \boldsymbol{x} \preceq \boldsymbol{t} - \frac{1}{\boldsymbol{\alpha}} \odot \mathbb{E}\left[ (\boldsymbol{t} - \boldsymbol{r}(\boldsymbol{p}(\boldsymbol{h}), \boldsymbol{h}))_+ \right]\\
& \left\Vert \mathbb{E}\left[ \boldsymbol{p} (\boldsymbol{h}) \right] \right\Vert_1 \leq P_0
\end{aligned}\, , \hspace{-8bp}
\label{eq:primal_riskaware_2}
\end{equation}
where ``$\odot$" stands for elementwise multiplication, while $(\cdot)_+$ and division with a vector are similarly overloaded.

Observe that problem \eqref{eq:primal_riskaware} is infinite-dimensional; in general such problems are challenging to tackle. Nonetheless, CV@R is a convex and monotone (in fact, coherent) risk measure \cite{Shapiro2014}, which preserves the convexity of both \eqref{eq:primal_riskaware} and \eqref{eq:primal_riskaware_2}. Therefore, under the effect of some appropriate constraint qualification, such as Slater's condition --assumed hereafter--, problems \eqref{eq:primal_riskaware} and \eqref{eq:primal_riskaware_2} exhibit no duality gap (in fact, strong duality). This fact suggests that we handle \eqref{eq:primal_riskaware_2} in the dual domain, i.e., within the framework of Lagrangian duality.


\section{Lagrangian Duality}
\label{subsec:dual_problem}

The Lagrangian of problem \eqref{eq:primal_riskaware_2} is defined as
\begin{equation}
\begin{aligned}
&\hspace{-8pt} \mathcal{L}(\boldsymbol{x},\boldsymbol{p}, \boldsymbol{t}, \Lambda, \mu)
\\
&\triangleq f_0(\boldsymbol{x}) + \mu \left(P_0 - \left\Vert \mathbb{E}\left[ \boldsymbol{p} (\boldsymbol{h}) \right] \right\Vert_1 \right)
\\
&\quad\,\,+ \Lambda^T \left[\boldsymbol{t} - \frac{1}{\boldsymbol{\alpha}} \odot \mathbb{E}\left[ (\boldsymbol{t} - \boldsymbol{r}(\boldsymbol{p}(\boldsymbol{h}), \boldsymbol{h}))_+ \right] - \boldsymbol{x} \right],
\end{aligned}
\end{equation}
where $\Lambda \succeq 0$ and $\mu \geq 0$ are the dual variables associated with the explicit constraints of \eqref{eq:primal_riskaware_2}. Accordingly, the \textit{dual function} is defined as the maximization of the Lagrangian function over the primal variable triplet $(\boldsymbol{x},\boldsymbol{p},\boldsymbol{t})$, i.e.,
\begin{equation}\label{dual_function}
D(\Lambda, \mu) = \sup_{\boldsymbol{x}\in \mathcal{X}, \boldsymbol{p} \succeq \boldsymbol{0}, \boldsymbol{t}}\ \mathcal{L}(\boldsymbol{x},\boldsymbol{p}, \boldsymbol{t}, \Lambda, \mu).
\end{equation}
Subsequently, the \textit{dual problem} is the minimization of the dual function over the dual variable pair $(\Lambda, \mu)$, i.e.,
\begin{equation}\label{dual}
\begin{aligned}
D^* &= \inf_{(\Lambda, \mu) \succeq \boldsymbol{0}}\ D(\Lambda, \mu),\\
&= \inf_{(\Lambda, \mu) \succeq \boldsymbol{0}}\ \sup_{\boldsymbol{x}\in \mathcal{X}, \boldsymbol{p} \succeq \boldsymbol{0}, \boldsymbol{t}}\ \mathcal{L}(\boldsymbol{x},\boldsymbol{p}, \boldsymbol{t}, \Lambda, \mu).
\end{aligned}
\end{equation}
As mentioned previously, problem \eqref{eq:primal_riskaware_2} exhibits strong duality (under Slater's condition), which means that $P^*=D^*$ and, what is more, optimal dual variables are guaranteed to exist. We also observe that even though problem \eqref{eq:primal_riskaware_2} is infinite-dimensional, its dual \eqref{dual} is finite-dimensional, which is a very useful fact if we are able to tackle the maximization involved in the dual function \eqref{dual_function} --in particular over $\boldsymbol{p}$-- adequately.

Leveraging strong duality, we hereafter focus on devising an efficient primal-dual algorithm for solving the minimax problem \eqref{dual}, hopefully providing an optimal solution to the constrained convex risk-aware problem \eqref{eq:primal_riskaware_2}, as well \cite{Ribeiro2012}.

\section{Risk-Aware Resource Allocation:\\ The Tail Waterfilling Algorithm}
\label{sec:resource_allocation}

The dual problem can be separated into several subproblems with respect to the primal variables. In particular, \eqref{dual} can be equivalently expressed as
\begin{equation}
\begin{aligned}\label{Dual_Open}
\inf_{(\Lambda, \mu) \succeq \boldsymbol{0} }\ 
\bigg\{ \mu P_0 + \sup_{\boldsymbol{x}\in \mathcal{X}}\ f_0(\boldsymbol{x}) - \Lambda^T \boldsymbol{x} +\sup_{\boldsymbol{t}}\ \sum_{i=1}^n \lambda_i t_i \\
\,\,+\, \mathbb{E}\left[ \sup_{p_i\ge0}\ -\mu p_i - \frac{\lambda_i}{\alpha_i} \big( t_i-r_i(p_i(h_i), h_i) \big)_+ \right] \hspace{-2pt}\bigg\},
\end{aligned}
\end{equation}
where interchanging the $\sup$ over $\boldsymbol{p}$ with expectation (integration) is justified in light of the \textit{interchangeability principle}; see, e.g., \cite[Theorem 7.92]{Shapiro2014}. This fact allows us to derive an optimal policy in closed-form.

\subsection{Optimal Resource Policy}

The particular policy subproblem for each user $i$ is
\begin{align}
\sup_{p_i \geq 0}\ -\mu p_i - \frac{\lambda_i}{\alpha_i} \left( t_i- \log \left( 1 + \frac{h_i p_i}{\sigma_i^2} \right) \right)_+.
\label{eq:p_subproblem}
\end{align}
The next result provides an optimal solution of subproblem \eqref{eq:p_subproblem}, determining the behavior of the optimal risk-aware policy $\boldsymbol{p}^* (\boldsymbol{h})$, parameterized by primal/dual variables $(\boldsymbol{t},\Lambda,\mu)$.
\vspace{2pt}
\begin{theorem}[Optimal Resource Policy]\label{Theorem}
An optimal solution to the $i$-th policy subproblem  \eqref{eq:p_subproblem} may be expressed as
\begin{equation}
\hspace{-1pt}\boxed{
p^*_i(h_i,\cdot)  \hspace{-1pt}=\hspace{-1pt} \begin{cases}
\left[ \dfrac{\lambda_i}{\mu \alpha_i} - \dfrac{\sigma_i^2}{h_i} \right]_+, & \text{if }\dfrac{\lambda_i}{\mu \alpha_i e^{t_i}} - \dfrac{\sigma_i^2}{h_i} < 0
\\ \vspace{-10pt}
\\
\dfrac{\sigma_i^2\left( e^{t_i} - 1 \right)}{h_i}, & \text{if } \dfrac{\lambda_i}{\mu \alpha_i e^{t_i}} - \dfrac{\sigma_i^2}{h_i} \geq 0
\end{cases}
}\,\hspace{0.5pt},\hspace{-1.5pt}
\label{eq:p_sol_full}
\end{equation}
whenever $\mu$ and $\lambda_i$ are not simultaneously zero; otherwise, it is optimal to choose $p^*_i(h_i)=0.$
\end{theorem}
\vspace{3pt} 

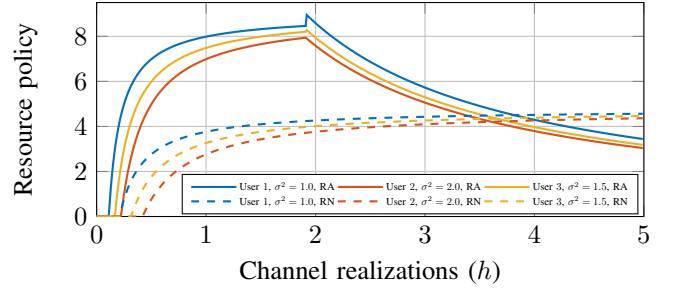
\begin{figure}[t]
\begin{tikzpicture}[trim axis right,baseline]
\begin{axis}[
    width=\linewidth,
    height=.5\linewidth,
    ylabel={Resource policy}, xlabel={Channel realizations ($h$)},
    xmin=0, xmax=5,
    ymin=0, ymax=9.5, ytick = {0, 2, 4, 6, 8},
    legend style={at={(0.99,0.01)},anchor=south east,
    nodes={scale=0.4, transform shape}},legend columns=3, grid]
\addplot[color=matblue, mark=no, thick] table[x=h,y=p_A_1] {Data/PA_RA_RN_plots.txt};
\addplot[color=matred, mark=no, thick] table[x=h,y=p_A_2] {Data/PA_RA_RN_plots.txt};
\addplot[color=matorange, mark=no, thick] table[x=h,y=p_A_3] {Data/PA_RA_RN_plots.txt};
\addplot[color=matblue, mark=no, dashed, thick] table[x=h,y=p_N_1] {Data/PA_RA_RN_plots.txt};
\addplot[color=matred, mark=no, dashed, thick] table[x=h,y=p_N_2] {Data/PA_RA_RN_plots.txt};
\addplot[color=matorange, mark=no, dashed, thick] table[x=h,y=p_N_3] {Data/PA_RA_RN_plots.txt};
\legend{ User $1 \text{, } \sigma^2=1.0 \text{, RA}$ , User $2 \text{, } \sigma^2=2.0 \text{, RA}$ , User $3 \text{, } \sigma^2=1.5 \text{, RA}$ , User $1 \text{, } \sigma^2=1.0 \text{, RN}$ , User $2 \text{, } \sigma^2=2.0 \text{, RN}$ , User $3 \text{, } \sigma^2=1.5 \text{, RN}$}
\end{axis}
\end{tikzpicture}
\caption{Optimal resource policies for risk-aware (RA, solid) and risk-neutral (RN, dashed) settings in a $3$-user network, with $\mu = 0.07$, $\alpha_i = 0.53$ and $\lambda_i = 0.33$ for all $i$, $t_1 = 2.9$, $t_2 = 2.15$ and $t_3 = 2.45$, respectively.}
\label{policy_plot}
\end{figure}

The risk-aware policy presented in Theorem \ref{Theorem} is a genuine extension of the classical risk-neutral waterfilling power policy \cite{Ribeiro2012, Cover2005}, which for user $i$ may be expressed as
\begin{equation}
p^N_i(h_i, \cdot)=\bigg[\dfrac{\lambda_{i}}{\mu}-\dfrac{\sigma_i^{2}}{h_i}\bigg]_{+}.
\end{equation}
This reduction is obtained by setting $\alpha_i=1$, and then sending the CV@R variable $t_i$ to $+\infty$ in \eqref{eq:p_sol_full} (this can be explained by the construction of the CV@R; see \cite[Chapter 6]{Shapiro2014}).

Presence of a confidence level $\alpha_i\in(0,1)$ (and the related rate target $t_i$) makes the policy \textit{less opportunistic} as compared with its risk-neutral counterpart. \textit{First}, power is allocated more aggressively to more faded channels (smaller values of $h_i$) in the $[\cdot]_+$-related part of the policy (cf. classical waterfilling), as $\alpha_i$ decreases and the corresponding branch of \eqref{eq:p_sol_full} is active (also depending on $t_i$). Further, if $\alpha_i$ (resp. $t_i$) is small enough to make the lower branch of \eqref{eq:p_sol_full} active,  constant relative to $\alpha$ but now proportional to $t_i$ power equal to $\sigma_i^2\left( e^{t_i} - 1 \right) / h_i$ is allocated. We observe a striking similarity of this term to outage-optimal policies; see, e.g., \cite{Li2005mac}. This might point to intricate information-theoretic properties of the CV@R approach, which can be the subject of future investigation.

\textit{Second}, when the $[\cdot]_+$-related part of the policy is nonpositive, then the policy indeed becomes opportunistic. However, given relevant (or optimized) values of $t_i$, the opportunism of the policy is effectively \textit{restricted to the lower tail} (occuring with proability equal to $\alpha_i$) of the random rate $r_i (p_i (h_i),h_i)$.

These remarks may also be readily observed in Fig. \ref{policy_plot}. Having explicitly determined $\boldsymbol{p}^*$, we may now derive primal --in $(\boldsymbol{x},\boldsymbol{t})$-- and dual --in $(\Lambda, \mu)$-- iterations, purposed to \textit{recursively learn} a globally optimal solution to \eqref{eq:primal_riskaware_2}. 

\subsection{CV@R Target / Risk-Ergodic Rate Updates--Primal Ascent}
\label{subsec:primal_estimation}

The remaining subproblems of \eqref{Dual_Open} relative to primal variables $(\boldsymbol{x},\boldsymbol{t})$ can be solved separately. Regarding maximization over $\boldsymbol{t}$, and with $p^*_i=p^*_i(h_i,t_i,\cdot)$, we end up with the problem
\begin{equation}
\sup_{t_i}\  \mathbb{E}\Bigg[ \lambda_i t_i - \mu  p^*_i-\frac{\lambda_i}{\alpha_i} \left( t_i-\log \left( 1 + \frac{h_i p^*_i}{\sigma_i^2} \right) \right)_+ \Bigg],
\label{eq:t_subproblem}
\end{equation}
for each $i$, where we explicitly denote the dependence of $p_i^*$ on $t_i$. It can be easily seen that the function inside the expectation of \eqref{eq:t_subproblem} is jointly concave relative to $(t_i,p_i)$. Therefore, it is also concave in $t_i$ under partial maximization over $p_i\ge0$.

By \cite[Theorem 7.52]{Shapiro2014}, it then follows that every subgradient of the latter is a stochastic subgradient of the objective of \eqref{eq:t_subproblem}. Due to initial joint concavity in $(t_i,p_i)$, it can be shown that such a stochastic subgradient may be selected as
\begin{align}
\hspace{-4pt}g_i(h_i, t_i,\cdot) \hspace{-1pt}=\hspace{-1pt} \lambda_i 
\hspace{-1pt}-\hspace{-1pt}
\frac{\lambda_i}{\alpha_i} H\left[t_i 
\hspace{-1pt}-\hspace{-1pt} 
\log\left( 1 \hspace{-1pt}+\hspace{-1pt} \frac{h_i p^*_i(h_i,t_i,\cdot)}{\sigma_i^2} \right)  \right] \hspace{-2pt}, \hspace{-4pt} \label{eq:t_subgradient}
\end{align}
where $H(\cdot)$ is the step (Heaviside) multifunction.
Then, provided an iteration index $n\in\mathbb{N}$ and processes $\{h_i^n\}$ and $\{t_i^{n}\}$ (and implicitly meant $\{\lambda_i^{n}\}$ and $\{\mu^{n}\}$; see below), we may formulate a stochastic subgradient ascent scheme for $t_i$ as
\begin{align}
t_i^n = t_i^{n-1} + \varepsilon_{t} g_i(h_i^n, t_i^{n-1},\cdot),\quad n\ge1, \label{eq:t_sub_ascent}
\end{align}
with stepsize $\varepsilon_t > 0$, and starting from some initial value $t^0_i$.

Maximization over $\boldsymbol{x}$, on the other hand, depends on the dual variables $\Lambda$ and concave utility $f_0$. Hereafter, we assume that an optimal solution as a function of $\Lambda$ --or $\Lambda^n,n\ge0$--
\begin{align}
\boldsymbol{x}^*(\Lambda) \in \arg\max_{\boldsymbol{x}\in \mathcal{X}}\ f_0(\boldsymbol{x}) - \Lambda^T \boldsymbol{x} \label{eq:x_general}
\end{align}
exists, and $f_0$ is such that $\boldsymbol{x}^*(\Lambda^n)$ is available (e.g., in closed form). Standard derivations and variable eliminations for popular utility functions, namely, sumrates and proportional fairness are provided later on, for completeness.

\subsection{Dual Variable Updates}
\label{subsec:dual_estimation}

Lastly, we can formulate stochastic (quasi-)subgradient descent updates for dual variables $(\Lambda,\mu)$. This is done along the lines of \cite{Ribeiro2012}, by exploiting the corresponding constraint gaps. Note that the dual function $D$ is convex and separable in $(\Lambda, \mu)$. For the power constraint multiplier $\mu$,  we have
\begin{align}
\mu^n = \left[ \mu^{n-1} - \varepsilon_{\mu} \left(P_0 - \sum_{i=1}^n p^*_i(h_i^n,\mu^{n-1},\cdot) \right) \right]_+, \label{eq:mu_grad_descent}
\end{align}
with stepsize $\varepsilon_\mu  > 0$, starting from $\mu^0$.
Similarly, for the rate constraint vector of multipliers $\Lambda$, we get, for each $i$,
\begin{align*}
\lambda^n_i =& \Bigg[ \lambda^{n-1}_i - \varepsilon_{\Lambda} \bigg(  -  x_i^*(\Lambda^{n-1}) + t^{n-1}_i  \numberthis \label{eq:lambda_grad_descent}\\
&- \frac{1}{\alpha_i}\left(t^{n-1}_i - \log\left(1 + \frac{h^{n}_i p^*_i(h_i^n,\lambda_i^{n-1},\cdot)}{\sigma_i^2} \right) \right)_+ \bigg) \Bigg]_+,
\end{align*}
with stepsize $\varepsilon_\Lambda > 0$, and starting from $\lambda_i^0$.

The complete description of the proposed primal-dual algorithm, which we suggestively call \textit{tail waterfilling}, is presented in Algorithm~\ref{alg:gen_alg}.

\subsection{Common Utilities}
\label{subsec:utilities}
\textit{Sumrate:} If $f_0(\boldsymbol{x}) = \boldsymbol{w}^T \boldsymbol{x}$, $\boldsymbol{w}\succeq \boldsymbol{0}$, $\boldsymbol{x}\in\mathbb{R}^n$, the  subproblem relative to $\boldsymbol{x}$ becomes $\sup_{\boldsymbol{x}}\ (\boldsymbol{w} - \Lambda)^T \boldsymbol{x}$,
which is unbounded for any $\boldsymbol{w}$ and $\Lambda$, except for the optimal dual choice $\Lambda = \boldsymbol{w}$. 
Of course, this step eliminates both variables $\boldsymbol{x}$ and $\Lambda$.

\begin{algorithm}[tbp]
\centering
\begin{algorithmic}
\STATE Choose initial values $\boldsymbol{t}^0, \boldsymbol{p}^0, \boldsymbol{x}^0, \mu^0, \Lambda^0$.
\FOR{$n = 1$ \textbf{to} Process End}
\STATE Observe $\boldsymbol{h}^n$.
\STATE \textit{\# Primal Variables}
\STATE $\boldsymbol{\to}$ Set $p_i^*(\cdot)$ using \eqref{eq:p_sol_full}, for all $i$.
\STATE $\boldsymbol{\to}$ Update $t^n_i$ using \eqref{eq:t_sub_ascent} and \eqref{eq:t_subgradient}, for all $i$.
\STATE $\boldsymbol{\to}$ Obtain $\boldsymbol{x}^*(\Lambda^{n-1})$ from \eqref{eq:x_general}.
\STATE \textit{\# Dual Variables}
\STATE $\boldsymbol{\to}$ Update $\mu^{n}$ using \eqref{eq:mu_grad_descent}.
\STATE $\boldsymbol{\to}$ Update $\lambda^n_i$ using \eqref{eq:lambda_grad_descent}, for all $i$.
\ENDFOR
\end{algorithmic}
\caption{Tail Waterfilling}
\label{alg:gen_alg}
\end{algorithm}

\textit{Proportional Fairness:} In this case, we can choose $f_0(\boldsymbol{x}) = \sum_{i=1}^n \log({x}_i)$, $\boldsymbol{x}\in\mathbb{R}^n$, and the subproblem in $\boldsymbol{x}$ becomes
\begin{equation}
\begin{aligned}
\sup_{\boldsymbol{x}}\ \sum_{i=1}^n \log(x_i) - \Lambda^T \boldsymbol{x} &= \sup_{\boldsymbol{x}}\ \sum_{i=1}^n \log(x_i) - \lambda_i x_i,
\end{aligned}
\label{eq:x_to_lambda}
\end{equation}
which gives $x^*_i = 1/\lambda_i$ for each $i$. 

\section{Performance Evaluation}
\label{sec:performance_eval}

Let us now verify and discuss the efficacy of the proposed tail waterfilling algorithm, summarized in Algorithm~\ref{alg:gen_alg}. We consider a basic $3$-terminal network consisting of Rayleigh fading point-to-point links with different noise variances. 
We then apply the tail waterfilling scheme for two utilities, namely, sumrate and proportional fairness. For the sumrate utility, the weights $\boldsymbol{w}$ are selected to average the individual risk-ergodic services $x_i$ per terminal, i.e., $w_i = 1/3, \forall i$. For both utilities, the stepsizes are set as $\varepsilon_t=10^{-3}$ and $\varepsilon_\mu=10^{-4}$, respectively. For proportional fairness, we set $\varepsilon_\Lambda=10^{-4}$. 

\begin{figure}[ht]
    \centering
    \begin{subfigure}[ht]{\linewidth}
    \begin{tikzpicture}[trim axis right,baseline]
    \begin{axis}[
        width=\linewidth,
        height=.4\linewidth,
        ylabel={Frequency}, 
        xmin=0, xmax=3, xmajorticks=false,
        ymin=0, ymax=0.19, ytick = {0, 0.1},
        legend image post style={scale=0.4},
        legend style={at={(0.99,0.99)},anchor=north east,
        nodes={scale=0.4, transform shape}}, grid]
    \addplot+[ybar interval,solid,black, very thin, fill=matblue, mark=no, ybar interval legend] table[x=rate,y=hist_1] {Data/a53_hist_outage_100e5.txt};
    \addplot+[ybar interval,solid,black, very thin, fill=matred, mark=no, ybar interval legend]  table[x=rate,y=hist_2] {Data/a53_hist_outage_100e5.txt};
    \addplot+[ybar interval,solid,black, very thin, fill=matorange, mark=no, ybar interval legend]  table[x=rate,y=hist_3] {Data/a53_hist_outage_100e5.txt};
    \addplot+[ybar interval,solid,gray, very thin, fill=matblue, mark=no, ybar interval legend] table[x=rate,y=hist_1] {Data/a100_hist_outage_100e5.txt};
    \addplot+[ybar interval,solid,gray, very thin, fill=matred, mark=no, ybar interval legend]  table[x=rate,y=hist_2] {Data/a100_hist_outage_100e5.txt};
    \addplot+[ybar interval,solid,gray, very thin, fill=matorange, mark=no, ybar interval legend] table[x=rate,y=hist_3] {Data/a100_hist_outage_100e5.txt};
    \legend{$\alpha=0.53 \text{, } \sigma^2 = 1.0$, $\alpha=0.53 \text{, } \sigma^2 = 2.0$, $\alpha=0.53 \text{, } \sigma^2 = 1.5$, $\alpha=1.00 \text{, } \sigma^2 = 1.0$, $\alpha=1.00 \text{, } \sigma^2 = 2.0$, $\alpha=1.00 \text{, } \sigma^2 = 1.5$}
    \end{axis}
    \end{tikzpicture}
    \end{subfigure}
    \\
    \begin{subfigure}[ht]{\linewidth}
    \begin{tikzpicture}[trim axis right,baseline]
    \begin{axis}[
        width=\linewidth,
        height=.4\linewidth,
        ylabel={Outage}, xlabel={Instantaneous Rates},
        xmin=0, xmax=3, grid,
        legend style={at={(0.99,0.02)},anchor=south east,
        nodes={scale=0.4, transform shape}}]
    \addplot[matblue, very thick] table[x=rate,y=out_1] {Data/a53_hist_outage_100e5.txt};
    \addplot[matred, very thick] table[x=rate,y=out_2] {Data/a53_hist_outage_100e5.txt};
    \addplot[matorange, very thick] table[x=rate,y=out_3] {Data/a53_hist_outage_100e5.txt};
    \addplot[matblue] table[x=rate,y=a100_1] {Data/data_outage_linear.txt};
    \addplot[matred] table[x=rate,y=a100_2] {Data/data_outage_linear.txt};
    \addplot[matorange] table[x=rate,y=a100_3] {Data/data_outage_linear.txt};
    \legend{$\alpha=0.53 \text{, } \sigma^2 = 1.0$, $\alpha=0.53 \text{, } \sigma^2 = 2.0$, $\alpha=0.53 \text{, } \sigma^2 = 1.5$, $\alpha=1.00 \text{, } \sigma^2 = 1.0$, $\alpha=1.00 \text{, } \sigma^2 = 2.0$, $\alpha=1.00 \text{, } \sigma^2 = 1.5$}
    \end{axis}
    \end{tikzpicture}
    \end{subfigure}
    \caption{Results for a network with $3$ users and a sum rate utility, $P_0 = 10$. Top: Rate histograms. Bottom: Outage probabilities.}
    \label{fig:rate_hist_linear}
    \vspace{-2.5pt}
\end{figure}
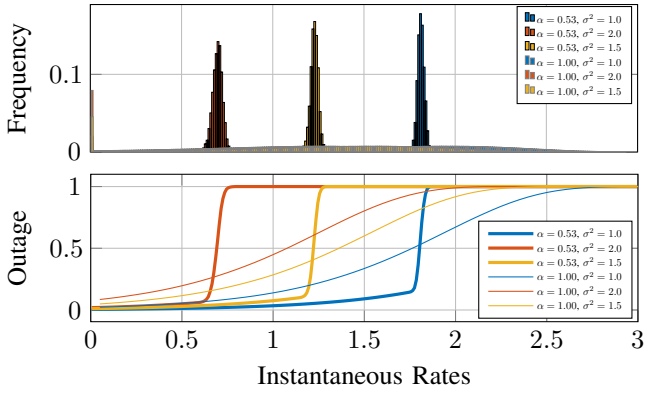

As shown in the histograms of Figs.~\ref{fig:rate_hist_linear} (top) and \ref{fig:rate_hist_pf} (top),
decreasing the (common) CVaR level $\alpha$ restricts the achievable rates in program \eqref{eq:primal_riskaware}, while constraining their volatility. This induces system robustness, since the system sustains a consistent and reliable level of performance, incurring infrequent rate drops in the long-term. Especially for proportional fairness, we observe that, \textit{simultaneously} with being robust, the rates are more fairly distributed among users with different noise levels. 

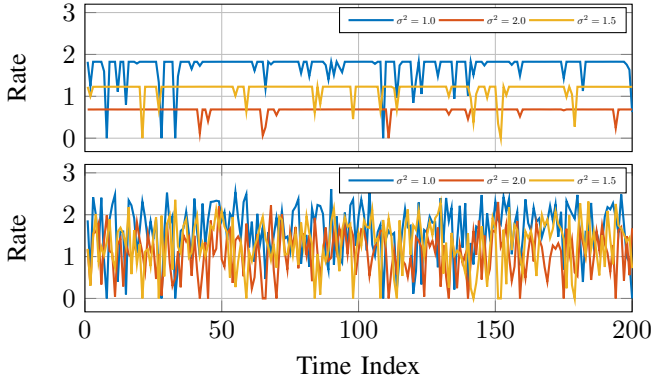
\begin{figure}[ht]
\centering
    \begin{subfigure}[ht]{\linewidth}
    \begin{tikzpicture}[trim axis right,baseline]
    \begin{axis}[
        width=\linewidth,
        height=.4\linewidth,
        ytick={0,1,2,3},
        ylabel={Rate},
        xmin=0, xmax=200, ymax=3.2, xmajorticks=false,
        legend style={at={(0.99,0.98)},anchor=north east,
        nodes={scale=0.4, transform shape}, legend columns = 3},
        grid]
    \addplot[matblue, thick, mark=no] table[x=time,y=val_1] {Data/a53_rate_vals_100e5.txt};
    \addplot[matred, thick, mark=no] table[x=time,y=val_2] {Data/a53_rate_vals_100e5.txt};
    \addplot[matorange, thick, mark=no] table[x=time,y=val_3] {Data/a53_rate_vals_100e5.txt};
    \legend{$\sigma^2 = 1.0$, $\sigma^2 = 2.0$, $\sigma^2 = 1.5$}
    \end{axis}
    \end{tikzpicture}
    \end{subfigure}
    \\
    \begin{subfigure}[ht]{\linewidth}
    \begin{tikzpicture}[trim axis right,baseline]
    \begin{axis}[
        width=\linewidth,
        height=.4\linewidth,
        xlabel={Time Index}, ylabel={Rate},
        xmin=0, xmax=200, ymax=3.2, 
        legend style={at={(0.99,0.98)},anchor=north east,
        nodes={scale=0.4, transform shape}, legend columns = 3},
        grid]
    \addplot[matblue, thick, mark=no] table[x=time,y=val_1] {Data/a100_rate_vals_100e5.txt};
    \addplot[matred, thick, mark=no] table[x=time,y=val_2] {Data/a100_rate_vals_100e5.txt};
    \addplot[matorange, thick, mark=no] table[x=time,y=val_3] {Data/a100_rate_vals_100e5.txt};
    \legend{$\sigma^2 = 1.0$, $\sigma^2 = 2.0$, $\sigma^2 = 1.5$}
    \end{axis}
    \end{tikzpicture}
    \end{subfigure}
    \caption{Achieved rates for the $3$-user network with a sum rate utility, $P_0 = 10$. Top: $\alpha = 0.53$ (risk-aware). Bottom: $\alpha = 1$ (risk-neutral).}
    \label{fig:rate_linear}
\end{figure}

\begin{figure}[ht]
    \centering
    \begin{subfigure}[ht]{\linewidth}
    \begin{tikzpicture}[trim axis right,baseline]
    \begin{axis}[
        width=\linewidth,
        height=.4\linewidth,
        ylabel={Frequency},
        xmin=0, xmax=3, xmajorticks=false,
        ymin=0, ymax=0.12,
        ytick = {0, 0.1},
        legend image post style={scale=0.4},
        legend style={at={(0.99,0.99)},anchor=north east,
        nodes={scale=0.4, transform shape}}, grid]
    \addplot+[ybar interval,solid,black, very thin, fill=matblue, mark=no, ybar interval legend] table[x=rate,y=hist1] {Data/rate_hist_out_pf_051.txt};
    \addplot+[ybar interval,solid,black, very thin, fill=matred, mark=no, ybar interval legend] table[x=rate,y=hist2] {Data/rate_hist_out_pf_051.txt};
    \addplot+[ybar interval,solid,black, very thin, fill=matorange, mark=no, ybar interval legend] table[x=rate,y=hist3] {Data/rate_hist_out_pf_051.txt};
    \addplot+[ybar interval,solid,gray, very thin, fill=matblue, mark=no, ybar interval legend] table[x=rate,y=hist1] {Data/rate_hist_out_pf_100.txt};
    \addplot+[ybar interval,solid,gray, very thin, fill=matred, mark=no, ybar interval legend] table[x=rate,y=hist2] {Data/rate_hist_out_pf_100.txt};
    \addplot+[ybar interval,solid,gray, very thin, fill=matorange, mark=no, ybar interval legend] table[x=rate,y=hist3] {Data/rate_hist_out_pf_100.txt};
    \legend{$\alpha=0.51 \text{, } \sigma^2 = 1.0$, $\alpha=0.51 \text{, } \sigma^2 = 2.0$, $\alpha=0.51 \text{, } \sigma^2 = 1.5$, $\alpha=1.00 \text{, } \sigma^2 = 1.0$, $\alpha=1.00 \text{, } \sigma^2 = 2.0$, $\alpha=1.00 \text{, } \sigma^2 = 1.5$}
    \end{axis}
    \end{tikzpicture}
    \end{subfigure}
    \\
    \begin{subfigure}[ht]{\linewidth}
    \begin{tikzpicture}[trim axis right,baseline]
    \begin{axis}[
        width=\linewidth,
        height=.4\linewidth,
        ylabel={Outage},  xlabel={Instantaneous Rates},
        xmin=0, xmax=3, grid,
        legend style={at={(0.99,0.02)},anchor=south east,
        nodes={scale=0.4, transform shape}}]
    \addplot[matblue, very thick] table[x=rate,y=out1] {Data/rate_hist_out_pf_051.txt};
    \addplot[matred, very thick] table[x=rate,y=out2] {Data/rate_hist_out_pf_051.txt};
    \addplot[matorange, very thick] table[x=rate,y=out3] {Data/rate_hist_out_pf_051.txt};
    \addplot[matblue] table[x=rate,y=out1] {Data/rate_hist_out_pf_100.txt};
    \addplot[matred]table[x=rate,y=out2] {Data/rate_hist_out_pf_100.txt};
    \addplot[matorange] table[x=rate,y=out3] {Data/rate_hist_out_pf_100.txt};
    \legend{$\alpha=0.51 \text{, } \sigma^2 = 1.0$, $\alpha=0.51 \text{, } \sigma^2 = 2.0$, $\alpha=0.51 \text{, } \sigma^2 = 1.5$, $\alpha=1.00 \text{, } \sigma^2 = 1.0$, $\alpha=1.00 \text{, } \sigma^2 = 2.0$, $\alpha=1.00 \text{, } \sigma^2 = 1.5$}
    \end{axis}
    \end{tikzpicture}
    \end{subfigure}
    \caption{Results for a network with $3$ users and a proportional fairness utility, $P_0 = 10$. Top: Rate histograms. Bottom: Outage probabilities. }
    \label{fig:rate_hist_pf}
    \vspace{-1.8pt}
\end{figure}
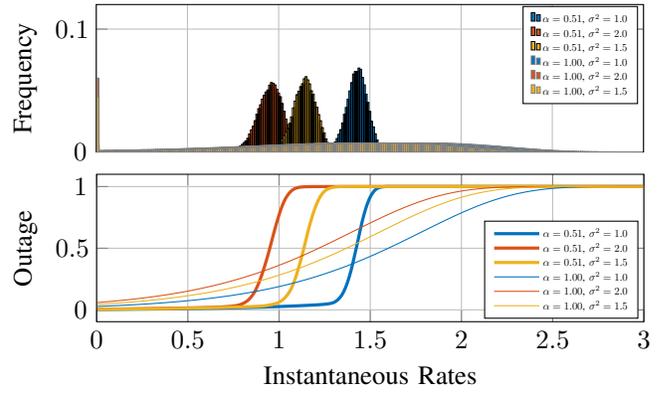
\begin{figure}[ht]
\centering
    \begin{subfigure}[ht]{\linewidth}
    \begin{tikzpicture}[trim axis right,baseline]
    \begin{axis}[
        width=\linewidth,
        height=.4\linewidth,
        ytick={0,1,2,3},
        ylabel={Rate}, xmajorticks=false,
        xmin=0, xmax=200, ymax=3.2, 
        legend style={at={(0.99,0.98)},anchor=north east,
        nodes={scale=0.4, transform shape}, legend columns = 3},
        grid]
    \addplot[matblue, thick, mark=no] table[x=time,y=val1] {Data/rate_vals_pf_051.txt};
    \addplot[matred, thick, mark=no] table[x=time,y=val2] {Data/rate_vals_pf_051.txt};
    \addplot[matorange, thick, mark=no] table[x=time,y=val3] {Data/rate_vals_pf_051.txt};
    \legend{$\sigma^2 = 1.0$, $\sigma^2 = 2.0$, $\sigma^2 = 1.5$}
    \end{axis}
    \end{tikzpicture}
    \end{subfigure}
    \\
    \begin{subfigure}[ht]{\linewidth}
    \begin{tikzpicture}[trim axis right,baseline]
    \begin{axis}[
        width=\linewidth,
        height=.4\linewidth,
        xlabel={Time Index}, ylabel={Rate},
        xmin=0, xmax=200, ymax=3.2, 
        legend style={at={(0.99,0.98)},anchor=north east,
        nodes={scale=0.4, transform shape}, legend columns = 3},
        grid]
    \addplot[matblue, thick, mark=no] table[x=time,y=val1] {Data/rate_vals_pf_100.txt};
    \addplot[matred, thick, mark=no] table[x=time,y=val2] {Data/rate_vals_pf_100.txt};
    \addplot[matorange, thick, mark=no] table[x=time,y=val3] {Data/rate_vals_pf_100.txt};
    \legend{$\sigma^2 = 1.0$, $\sigma^2 = 2.0$, $\sigma^2 = 1.5$}
    \end{axis}
    \end{tikzpicture}
    \end{subfigure}
    \caption{Achieved rates for the  $3$-user network with a proportional fairness utility, $P_0 = 10$. Top: $\alpha = 0.51$ (risk-aware). Bottom: $\alpha = 1$ (risk-neutral).}
    \label{fig:rate_pf}
\end{figure}
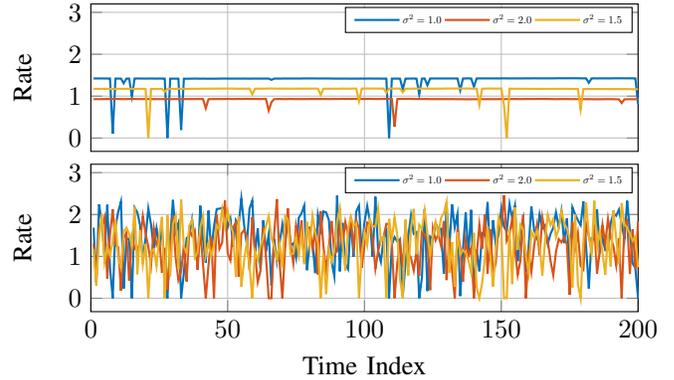

Equivalent remarks are in order regarding Figs.~\ref{fig:rate_hist_linear} (bottom) and \ref{fig:rate_hist_pf} (bottom), which show 
the user outage probabilities, i.e., the cumulative distribution function $P_{\mathrm{out}}(r_o) = P\{r \leq r_o \}$, as another intuitive measure to evaluate system robustness. 
In fact, we observe that for lower values of $\alpha$, the system exhibits
\textit{lower and sharper} probabilities of outage --optimally tuned in the CV@R sense-- at \textit{always attainable} channel rates. On the other hand, less risk-aware settings corresponding to higher values for $\alpha$ result in much higher variability in the corresponding optimal channel rates.

The latter observations are also evident from Figs.~\ref{fig:rate_linear} and \ref{fig:rate_pf}, which highlight the vast difference in rate variability between the risk-aware and risk-neutral policies, through their evolution in time (channel use). We see that the optimal CV@R policy exhibits quasi-invariant communication rate trends,
keeping the rates at certain reliability levels. 
Further, the proportional fairness utility achieves more evenly distributed rates among users, as shown in Fig.~\ref{fig:rate_pf} (top). In other words, the combination of risk-awareness and proportional fairness achieves system performance that is \textit{both} user-fair, \textit{and} aware of fading risk.

\section{Conclusion}
\label{sec:conclusion}

We proposed a new risk-aware reformulation of the classical resource allocation problem for point-to-point networks. Utilizing the CV@R as a measure of risk generalizing expectations, we developed the tail waterfilling algorithm, which extends classical stochastic waterfilling in an interpretable and tunable way, and  induces network robustness and reliability rigorously and tractably. 
The effectiveness of tail waterfilling was corroborated via detailed numerical simulations.

\bibliographystyle{IEEEtran}
\bibliography{Files/references}

\end{document}